\magnification=\magstep1
\tolerance 500
\bigskip
\centerline{\bf Lee-Friedrichs Model}
\bigskip
\centerline{Lawrence P. Horwitz}
\centerline{\it School of Physics}
\centerline{\it Raymond and Beverly Sackler Faculty of Exact Sciences}
\centerline{\it Tel Aviv University, Ramat Aviv, Israel}
\centerline{\it and}
\centerline{\it Department of Physics}
\centerline{\it Bar Ilan University}
\centerline{\it Ramat Gan, Israel}
\bigskip
\par What is known today as the Lee-Friedrichs model$^{1,2}$ is
characterized by a self-adjoint operator $H$ on a Hilbert space
${\cal H}$, which is the sum of two self-adjoint operators $H_0$ and
$V$, such that $H,H_0$ and $V$ have common domain; $H_0$ has
absolutely continuous spectrum (of uniform multiplicity)  except for
 the end-point of the semi-bounded from below spectrum, and one or
 more eigenvalues which may or may not be embedded in the continuum.
The operator $V$ is compact and of finite rank, and induces a map from
the subspace of ${\cal H}$ spanned by the eigenvectors of $H_0$ to the
subspace corresponding to the continuous spectrum (and the reverse).
The central idea of the model is that $V$ does not map the subspace
corresponding to the continuous spectrum into itself, and, as a
consequence, the model becomes solvable in the sense that we shall
describe below.
\par In the physical applications of the model, $H$ corresponds to the
Hamiltonian operator, the self-adjoint operator (often the
self-adjoint completion of an essentially self-adjoint operator) that
generates the unitary evolution (through Schr\"odinger's equation) of
the vector in ${\cal H}$ representing the state of the physical
system in time.
\par The resolvent $G(z) = (z-H)^{-1}$ generated by the Laplace transform
on $[0,\infty)$ by $e^{izt}$  on the Schr\"odinger evolution operator
$e^{-iHt}$ (both acting on some suitable $f \in {\cal H}$) is
analytic in the upper half $z$-plane.  Denoting by $\langle
\lambda \vert f)$ (with Lebesgue measure $d\lambda$) the
representation of $f \in {\cal H}$ on the continuous spectrum
$\lambda$ of $H_0$ on $[0,\infty)$ and $\phi \in {\cal H}$ the
eigenvector with eigenvalue $E_0$ (assuming for this illustration
just one discrete eigenvector), we see that the second resolvent
equation
$$ G(z)= G_0(z) +G_0(z) V G(z), \eqno(1)$$
where $G_0(z) = (z-H_0)^{-1}$, can be exactly solved by the pair of
equations
$$(\phi, G(z)\phi) = {1 \over {z-E_0}} + {1 \over
{z-E_0}}\int_0^\infty \,d\lambda \,(V\phi\vert \lambda \rangle
\langle \lambda \vert G(z) \phi)
\eqno(2)$$
 and
$$ \langle \lambda \vert G(z)\phi) = {1 \over z-\lambda}\langle
\lambda \vert V\phi) (\phi, G(z)\phi). \eqno(3)$$
Substituting $(3)$ into $(2)$, we see that
$$ h(z) (\phi, G(z)\phi) \equiv \bigl( z-E_0 - \int_0^\infty {\vert
(V\phi\vert \lambda \rangle \vert^2 \over z-\lambda} d\lambda \bigr)
(\phi, G(z) \phi) = 1. \eqno(4)$$
If the discrete spectral value $E_0$ is separated from the continuum
($E_0 <0$), then $(\phi, G(z)\phi)$ has a pole on the real axis at the
point
$$ E_1 = E_0 + \int_0^\infty {\vert (V\phi \vert \lambda \rangle
\vert^2 \over E_1 - \lambda} d\lambda  <0. \eqno(5)$$
If $E_0$ is embedded in the continuum that lies on $[0,\infty)\,\,(E_0
>0)$, one can avoid the generation of a real pole on the negative half
line by the inequality
$$ \int_0^\infty {\vert (V\phi \vert \lambda \rangle \vert^2 \over
\lambda } d\lambda <E_0. $$
The projection of the time evolution of the quantum mechanical state
represented by $\phi$ back onto the initial state is given by (we use
units in which $\hbar$, the Plack constant divided by $2\pi$, is unity)
$$ (\phi, e^{-iHt}\phi) = {1 \over 2\pi i} \int_{\atop C}
e^{-izt}(\phi, G(z)\phi) dz, \eqno(6)$$
where the contour goes from $+\infty$ in the negative direction of the
real axis and a small distance above it, around the branch point
counterclockwise at $0$, and back to $+\infty$ below the real axis.
The construction defined on the left hand side of
 $(6)$ was used by Wigner and Wiesskopf$^3$ in 1930 as
a model for the description of unstable systems; they used it to
calculate the linewidth of a radiating atom.
\par The contour of the integral in Eq. $(6)$
 can be deformed so that the integration below the real
line is shifted to the negative imaginary axis where, for t
sufficiently positive, this contribution can be considered as
negligible (except near the branch cut).  The integral path above the
real axis can be similarly deformed into the second Riemann sheet of the
function $h(z)^{-1}$ to the negative real axis, but there is a
possibility that the second sheet extension of this function has a
pole in the lower half plane.  We see this in the Lee-Friedrichs
model$^{1,2}$ by studying (for $\zeta$ real)
$$ \eqalign{h(\zeta +i\epsilon) - h(\zeta-i\epsilon) &= \int_0^\infty
\vert(V\phi\vert \lambda\rangle\vert^2 \bigl({1 \over {\zeta - \lambda -
i\epsilon}} - {1 \over {\zeta - \lambda + i\epsilon}} \bigr) d\lambda
\cr
&= 2\pi i \vert(V\phi \vert \zeta>\vert^2 .\cr}$$
Choosing a $V$ so that $W(\zeta) = \vert (V\phi \vert \zeta \rangle
\vert^2$ is the boundary value on the real axis of an analytic
function in some (sufficiently large) domain in the lower half plane,
we see that the second sheet continuation of $h(z)$ is
$$ h^{II}(z) = h(z) + 2\pi i W(z). \eqno(7)$$
Now,
$$ {\rm Im} h^{II}(z) = {\rm Im} z \bigl( \int_0^\infty
{\vert (V\phi \vert
\lambda \rangle \vert^2 \over \vert z -\lambda \vert^2 }d\lambda \bigr)  + 2
\pi {\rm Re} W(z);$$
if the value of $z$ that we seek is sufficiently close to the real
axis, ${\rm Re} W(z) >0$, and ${\rm Im} h^{II}(z)$ may have a zero
for  ${\rm Im} z <0$. If the real part vanishes as
well, one has a pole of $h^{II}(z)^{-1}$ at, say ${\hat z}$, which
implies a decay law of the time evolution of the so-called survival
amplitude $(6)$ of the form $e^{-i{\hat z}t}$, an exponential decay.
The imaginary part of ${\hat z}$ is the semi-decay width computed in
lowest order perturbation theory by Wigner and Weisskopf.$^3$
\par In quantum mechanical scattering theory$^4$, the scattered wave
is expressed in terms of an operator valued function of $z$
$$ T(z) = V + VG(z) V, \eqno(8)$$
analytic in the same domain as $G(z)$.
The transition amplitude $\langle \lambda \vert T(z)
 \vert \lambda' \rangle$
contains, by the hypotheses of the Lee-Friedrichs model, the reduced
resolvent $(\phi, G(z) \phi)$, and the second sheet pole discussed
above dominates the behavior of the scattering for an interval of
energies near the real part of ${\hat z}$, appearing as a scattering
resonance.  Hence the Lie-Friedrichs model offers an opportunity to
describe scattering and resonance phenomena, along with the behavior
of an unstable system, in the framework of a single mathematical
model$^5$.
\par The pole in $h^{II}(z)$ at ${\hat z}$ suggests that in some sense,
there may be an eigenvalue equation of the form
$$ z f(z) = Hf(z). \eqno (9)$$
 This equation is exactly solvable in the Lee-Friedrichs model, with
$$ \langle \lambda \vert f(z)) = {1 \over \lambda - z} \langle \lambda
\vert V \phi) (\phi, f(z)), \eqno(10)$$
but the eigenvalue equation $(9)$ is satisfied only after analytic
continuation to ${\hat z}$ in the same way as described above.  This
analytic continuation can be done in terms of the sesquilinear form
$(g, f(z))$ for a suitable $g \in {\cal D} \subset {\cal H}$, such
that $\langle \lambda \vert g)$ is the boundary value of an analytic
function on an adequate domain in the lower half plane (including the
point ${\hat z}$ within its boundary). [The eigenfunction $\phi$ must lie
 in ${\cal D}$ as well.]
  The Banach space functional
$f$ defined in this way lies in the space ${\overline {\cal H}}$ dual to
${\cal D}$, for which $ {\overline {\cal H}} \supset {\cal H} \supset
{\cal D}$, i.e., an element of a Gel'fand triple$ ^{6,7}$.  This
construction has provided the basis for useful physical applications.$^8$
 \par We remark that the quantity $(\phi, e^{-iHt} \phi)$ studied in
the Wigner-Weisskopf theory$^3$ can never be precisely exponential in
form (i.e., more generally,$Pe^{-iHt}P$, where $P$ is a projection,
cannot  be a semigroup)$^9$, although for sufficiently large (but not
too large) $t$, it may well approximate an exponential.  For example,
the $t$-derivative of $\vert (\phi, e^{-iHt}\phi)\vert^2$ at $t=0$
vanishes if $H\phi$ is defined. The time dependence of the
 Gel'fand triple function may, however,
be exactly exponential ( if ${\cal D}$ is sufficiently stable).
 \par The original model of Lee$^1$, formulated in the framework of
non-relativistic quantum field theory, was motivated by an interest in
the process of renormalization; it can be seen from $(5)$ that the
interaction V induces a shift in the point spectrum. There
is a conserved quantum number in Lee's field theory which enables the
model to be written in one sector as a quantum mechanical model
equivalent to the structure used by Friedrichs$^2$, whose motivation
was to study the general framework of the perturbation of continuous
spectra.
\par A relativistically covariant form of the Lee-Friedrichs model has been
 developed in ref. 10.
\bigskip
\noindent
REFERENCES
\smallskip
\frenchspacing
\item{1.} T.D. Lee, {\it Phys. Rev} {\bf 95} (1956) 1329.
\item{2.} K.O. Friedrichs, {\it Comm. Pure Appl. Math.} {\bf 1} (1948)
 361.
\item{3.} V.F. Weisskopf and E.P. Wigner, {\it Zeits. f. Phys.} {\bf 63}
(1930) 54; {\bf 65} (1930) 18.
\item{4.} For example, J.R. Taylor, {\it Scattering Theory}, John Wiley
and Sons, N.Y. (1972); R.J. Newton, {\it Scattering Theory of
Particles and Waves}, McGraw Hill, N.Y. (1976).
\item{5.} L.P. Horwitz and J.-P. Marchand {\it Rocky Mtn. J. Math. }
{\bf 1} (1971) 225.
\item{6.} I.M. Gel'fand and G.E. Shilov, {\it Generalized Functions},
Vol. 4, Academic Press, N.Y. (1968).
 \item{7.}  W. Baumgartel, {\it Math. Nachr.} {\bf 75} (1978) 133;
 T. Bailey and W.C. Schieve,
{\it Nuovo Cim.} {\bf 47A} (1978) 231;  A. Bohm, {\it The Rigged Hilbert
Space and Quantum Mechanics, Springer Lecture Notes on Physics} {\bf
78}, Springer, Berlin (1978);  L.P. Horwitz and I.M. Sigal, {\it Helv.
Phys. Acta} {\bf 51} (1980) 685;
 G. Parravicini, V. Gorini and E.C. G. Sudarshan, {\it Jour.
Math. Phys.} {\bf 21} (1980) 2208; A. Bohm, M. Gadella and G.B. Mainland
{\it Amer. J. Phys.} {\bf 57} (1989) 1103; A. Bohm and M. Gadella, {\it Dirac
Kets, Gamow Vectors and Gel'fand Triples, Springer Lecture Notes in
 Physics} {\bf 348} Springer Verlag, N.Y. (1989). .
\item{8.}  D. Cocolicchio, {\it Phys. Rev} {\bf 57}
(1998) 7251.  See also A. Grecos and I. Prigogine,{\it Physica} {\bf 59} (1972)
77; {\it ibid} {\it Proc. Nat. Aca. Sci. USA} {\bf 60} (1972) 1629;
T. Petroski, I. Prigogine and S. Tasaki, {\it Physica A} {\bf 175}(1991) 175;
I. Antoniou and I. Prigogine {\it Physica A} {\bf 192} (1993) 443; I.
 Antoniou and S. Tasaki, {\it Int. Jour. Quantum Chem.} {\bf 44} (1993) 425;
I. Antoniou, L. Dmitrieva, Y. Kuperin and Y. Melnikov, {\it Comp. Math.
 Appl.} {\bf 34} (1997) 425.
\item{9.} L.P. Horwitz, J.-P. Marchand and J. LaVita, {\it Jour.
Math. Phys.} {\bf 12} (1971) 2537; D. Williams, {\it Comm. Math. Phys}
 {\bf 21} (1971) 314.
\item{10.} L.P. Horwitz, {\it Found. Phys.} {\bf 25} (1995) 39.  This work is
discussed briefly in the first of ref. 8, Section III.
  See also C.J. Hammer and T.A.
 Weber, {\it Phys. Rev.} {\bf D5} (1972) 3087; D.N. Williams, {\it Nuclear
Phys.} {\bf B264} (1986)423; I. Antoniou, M. Gadella, I. Prigogine
and G.P. Pronko, {\it Jour. Math. Phys.} {\bf 39} (1998) 2995.

\vfill
\eject
\end
\bye